# A novel control mode of bionic morphing tail based on deep reinforcement learning

Liming Zheng[1], Zhou Zhou[2], Pengbo Sun[3], Zhilin Zhang[3], and Rui Wang[3]

*Abstract*—In the field of fixed wing aircraft, many morphing technologies have been applied to the wing, such as adaptive airfoil, variable span aircraft, variable swept angle aircraft, etc., but few are aimed at the tail. The traditional fixed wing tail includes horizontal and vertical tail. Inspired by the bird tail, this paper will introduce a new bionic tail. The tail has a novel control mode, which has multiple control variables. Compared with the traditional fixed wing tail, it adds the area control and rotation control around the longitudinal symmetry axis, so it can control the pitch and yaw of the aircraft at the same time. When the area of the tail changes, the maneuverability and stability of the aircraft can be changed, and the aerodynamic efficiency of the aircraft can also be improved. The aircraft with morphing ability is often difficult to establish accurate mathematical model, because the model has a strong nonlinear, model-based control method is difficult to deal with the strong nonlinear aircraft. In recent years, with the rapid development of artificial intelligence technology, learning based control methods are also brilliant, in which the deep reinforcement learning algorithm can be a good solution to the control object which is difficult to establish model. In this paper, the model-free control algorithm PPO is used to control the tail, and the traditional PID is used to control the aileron and throttle. After training in simulation, the tail shows excellent attitude control ability.

*Index Terms*-- Bionic Morphing Tail; Deep Reinforcement Learning; Attitude Control; Application; Dynamics Analysis.

## I. INTRODUCTION

In recent years, UAV has been widely used in agriculture, security, environmental protection, photography and other fields [1]. With the development of UAVs, people have higher requirements for their functions. They hope that a UAV can be competent for some complex tasks. The morphing UAV has gradually become a research hotspot around the world. In the field of multi-rotors, foldable UAVs [2] have emerged, which can cross obstacles in a smaller form. There are also multi powered dragon shaped multi rotors [3]. In the field of fixed wing, Slivestro et al. [4] summarized the technology of deformable aircraft used in history, including the aircraft with variable wingspan, the aircraft that can change the airfoil, and the aircraft that can switch the state of multi-models. Bionic morphing aircraft has been developed in recent years. In different environments and different flight stages, aircraft with appropriate shape can better balance the mobility, stability, aerodynamic efficiency and flight speed. The advantages of these deformations can be found in birds [5]. Inspired by bird flying, scholars around the world have made a series of morphing aircrafts to achieve certain performance improvements. For example, the morphing wings can improve the robustness of flight in the wind. Inspired by the feathers of birds, the wing is designed as a feather wing with variable area, so as to expand the flight envelope of the aircraft, and has better wind resistance performance. In addition, the imbalance of left and right lift caused by the area difference between the left and right wings can be used to control the aircraft roll [6]. In addition, a series of active deformation wings were designed to improve the aerodynamic efficiency [7].

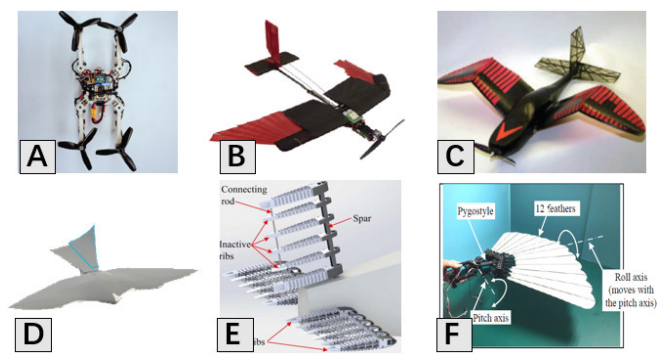

Figure 1. Example of other morphing wings and tails. (A) Morphing drone [2]. (B) Morphing wing [6]. (C) Bioinspired wing [7]. (D) MFC tail [8]. (E) SMA tail [9]. (F) Feather tail [10]

The wing morphing technology has a good balance of many performances of the aircraft, but few scholars have studied the morphing technology for the tail. In the stage of aircraft design, we know that different area of tail has different static stabilities, and big tail mean better control efficiency and bad aerodynamic efficiency. Through the integrated application of new intelligent materials, there are also some new tails of aircrafts. Lawren et al. [8] designed a kind of tail with MFC strain material, which can change the shape of the tail by changing the voltage to control the pitch and yaw of the aircraft. There are also flat tail and vertical tail based on SMA to change the camber of the airfoil [9]. In the use of this new material, the airfoil can be deflected by 10.7 degrees. In natural, the tail can ensure the stability of birds. When the flight phase requires high mobility and stability, such as hunting, fighting, landing and so on, birds will open the tail feathers to obtain better stability and control performance. For the bionic tail, Francis et al. designed a feather type tail, which can open and draw back the feathers, and control the area change of the tail by changing the overlapping area between the feathers [10], like a real tail of bird. However, [10] only studies the opening

*Research supported by the Seed Foundation of Innovation and Creation for Graduate Students in Northwestern Polytechnical University.

[1]Liming Zheng is with the School of Aeronautics, Northwestern Polytechnical University, (e-mail: 15719283618@163.com ).

[2]Zhou Zhou is with the School of Aeronautics, Northwestern Polytechnical University (corresponding author): Phone: +86-15719283618 e-mail: zhouzhou@nwpu.edu.cn.

[3]Pengbo Sun, Zhilin Zhang and Rui Wang are the School of Aeronautics, Northwestern Polytechnical University.

and recovery mechanism of feathers, and does not study its aerodynamic efficiency and control.

At present, the research on intelligent morphing aircraft focuses on the realization of deformable materials [11] and some structure designs, but the research on intelligent flight control method of intelligent morphing aircraft is relatively insufficient [12]. How to solve the control problem of morphing aircraft is a difficult problem, because the morphing aircraft often has a strong nonlinear mathematical model. Secondly, how to reasonably optimize the control strategy of the aircraft to adapt to different flight missions is another problem. Recently, Google DeepMind team proposed deep reinforcement learning [13], which optimizes the control strategy through the continuous interaction between the agent and the environment to obtain the maximum reward. In addition, reinforcement learning can be used for model-free control, which is undoubtedly a clever control method for aircraft which is difficult to model. These algorithms have achieved good results in go and other competitions. With the development of DRL, this kind of algorithm can have the output of continuous action, so scholars gradually apply the algorithm to the control of aircraft. For the deformed aircraft, Dan et al. [12] designed a feather morphing wing, which can change the wing area, and uses DDPG [14] algorithm to control the wing area to adapt to different environments. Valasek et al. [15] uses reinforcement learning to optimize the control system according to the aerodynamic and structural characteristics of the aircraft to obtain better aerodynamic efficiency. At the same time, more and more colleges and universities around the world also began to use DRL algorithm to control aircrafts. Reddy et al. [16] uses reinforcement learning algorithm to obtain good gliding in the field, and the aircraft can fly in the thermal for a longer time. For multi rotor attitude control, William et al. [17] compared the control results of traditional PID and DRL algorithms PPO [18], TRPO and DDPG. These DRL algorithms have the same good effect as PID in terms of rise time, stability and overshoot, and some even surpass PID. Also, for the attitude control of fixed wing, Bohn et al. [19] used PPO algorithm, and also achieved good control result, which exceeded the PID performance in the evaluation criteria such as rise time.

*A. Contributions*

Inspired by the structure and control method of bird's tail, this paper will show a feather type morphing tail, which is composed of multiple feathers, similar to [6] to control the effective area of tail by changing the overlapping area between feathers. Different from the traditional tail, the morphing tail has three dimensions of control, including deflection control $\delta_e$, area control $\delta_s$ and rotation control $\delta_r$ around the longitudinal symmetry axis. Therefore, the tail can have the tail control mode similar to that of birds, and can control the pitch and yaw of the aircraft at the same time. Different from [10], this morphing tail considers aerodynamic optimization and structural design, and shows certain advantages in dynamics. By changing the area, the stability, control efficiency and aerodynamic efficiency of the aircraft can be improved. Combined with different $\delta_e$ and $\delta_r$, the morphing tail has different control effect on pitch and yaw moment of aircraft, and the results are shown in Table 1.

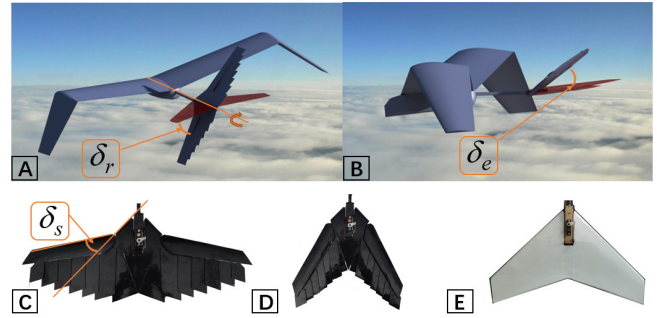

Figure 2. The morphing tail with 3 control variables. (A) The rotation control $\delta_r$ around the longitudinal symmetry axis (according to the right-hand rule, the rotation around the body axis x is positive). (B) The deflection control $\delta_e$ (downward deflection is positive). (C) The area control $\delta_s$, which is controled by the sweep angle of tail (forward is positive relative to the fixed part, namely, increasing area of tail). C is the largest area of the tail. (D) This is the smallest area of tail. (E) A normal tail

TABLE I. PITCH AND YAW CONTROL EFFECT OF TAIL

| Direction of $\delta_e$ | Direction of $\delta_r$ | Pitch moment M | Yaw moment N |
|---|---|---|---|
| + | + | − | + |
| + | − | − | − |
| − | + | + | − |
| − | − | + | + |

When the area of the morphing tail changes and rotates around the longitudinal symmetry axis of the aircraft, the dynamic model of the aircraft has obvious nonlinear changes. Therefore, this paper attempts to use model-free deep reinforcement learning control algorithm PPO and the traditional PID control algorithm to control the aircraft attitude, in which PPO controls the three control variables of the tail, and PID controls the aileron of the aircraft. Through the flight data online training control strategy, the final completed controller can control the attitude of aircraft very well. This paper will show the excellent performance of the hybrid control algorithm of PPO and PID through several criterions.

*B. Structure of the Paper*

The remainder of this letter is organized as follows. In Sec. Ⅱ we show the design of bionic tail, including mechanical design and pneumatic design. In Sec. Ⅲ we introduce the dynamic modeling of aircraft. In Sec. Ⅳ we introduce the control algorithm of the tail. In Sec. Ⅴ we show the model training and the result in the simulation environment. In Sec. Ⅵ we draw the conclusion and further research plan.

II. MORPHING TAIL DESIGN

This section will introduce the mechanical and aerodynamic design of the bionic tail, as well as its dynamic analysis.

*A. Mechanical Design*

The structural strength and mass are taken into account in the design of the tail wing to ensure that the deformation does not bring too much structural weight cost. After the airfoil design, it is necessary to ensure that the surface is smooth as far as possible, so as to reduce the aerodynamic drag. In this paper, the morphing tail is designed based on a traditional

UAV. In order to verify the feasibility and reliability of the design, we made a real morphing tail.

The mechanism with variable area adopts the deformation principle of parallelogram. If one angle of the parallelogram is changed with the servo, all the angles of the parallelogram will change, and then the overlapping area between feathers can be changed. In order to achieve better control effect, we should try our best to ensure that the proportion of the area of A (Figure 3) and C (Figure 3) is larger, that is, the area with variable area is expected to be larger, so as to improve the effect of deformation.

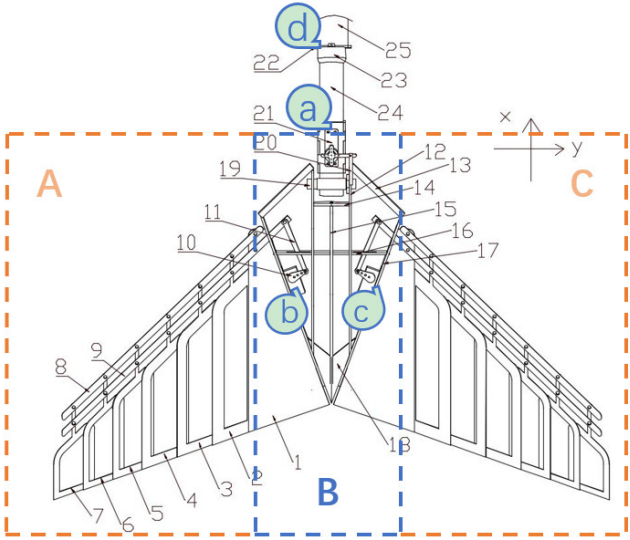

Figure 3. Mechanical design of the morphing tail.

The Figure 3 shows the schematic diagram of the internal structure of the morphing tail (excluding the rectification mechanism of the leading edge of the tail). (a) is the servo controlling the $\delta_e$. (b) and (c) are the servos controlling $\delta_s$, and the control signals of them are the same, namely, the angle changing synchronously. (d) is the motor that controls $\delta_r$. Marks (1) ~ (7) are feathers, among which (1) is a fixed feather and (2) ~ (7) are the movable feathers. (8) and (9) are the long sides of the parallelogram. (10) and (11) are the steering wheel and steering linkage respectively, used to conduct the servo angle control. (12) ~ (18) are the structural design of the fixed part of the morphing tail. (19) is the shaft controlling the $\delta_e$, (20) and (21) are used to fix the servo (a). (24) is the fixed carbon structure of the tail, (25) is the part of fuselage, (22) and (23) are used to connect the tail and fuselage. In order to reduce the weight of the feather, carbon fiber with large structural stiffness and low density is selected as the feather material, and light skin is selected for the feather surface, and 0.2mm carbon fiber sandwich balsa wood composite is selected for the frame structure of feather. The structure of the fixed part of the morphing tail is composed of different woods. The final mass of the morphing tail is 134g, which is only 28g higher than 106g of the traditional tail. It is a small amount compared with the mass of 1.14kg of the whole aircraft. Therefore, the mass cost of using the morphing tail is very small.

According to the above mechanical structure design, the final area variation range of the tail is shown in Figure 4. Due to the limitation of the structural form, the acceptable variation range is (-20,30), and the area changes from $0.0560 m^2$ to $0.0838 m^2$, with an area increase of 49.6%.

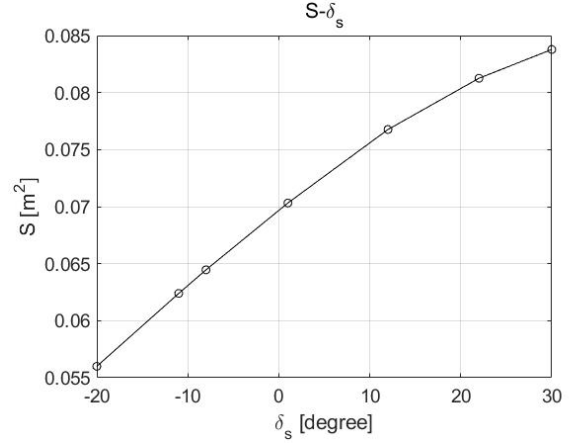

Figure 4. Area of tail changed by $\delta_s$

TABLE II. BOUNDARY OF TAIL AREA

| | $\delta_s$ [degree] | Tail area [$m^2$] |
|---|---|---|
| Minimum area | -20 | 0.0560 |
| General status | 0 | 0.0684 |
| Maximum area | 30 | 0.0838 |

### B. Aerodynamic Design

The feathers of birds are of sheet structure. The mechanical feathers need certain control equipment and some other structures, so the surface of the tail is not smooth enough and brings certain aerodynamic drag. However, the problem can be solved by designing the airfoil as shown in Figure 5. The airfoil is improved on the basis of NACA0010. The $l_{cover}$ is used to balance the aerodynamic performance of the tail, the covering effect on the mechanical structure and the area of morphing part. With the decrease of $l_{cover}$, the aerodynamic drag increases, and the covering effect becomes worse, but the area of the variable region becomes larger. Therefore, the chord strength of 37.5% is taken as a balance.

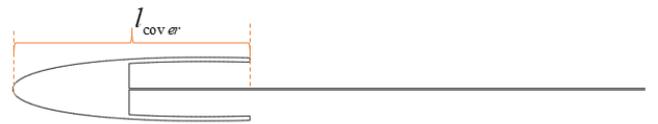

Figure 5. Airfoil of the morphing tail (based on the NACA0010)

### C. Dynamics Analysis

It has obvious advantages in dynamics, such as stability, aerodynamic efficiency and control efficiency. The change of the morphing tail area will bring about the change of the aircraft center of gravity position, as well as the change of the aerodynamic center position of the aircraft. When the morphing tail area becomes smaller, the sweep angle of the tail becomes larger, and the center of gravity moves slightly backward. At the same time, due to the reduction of the area of the tail, the aerodynamic center of the whole aircraft will move forward, and the center of gravity and aerodynamic center of the aircraft will gradually approach, resulting in the reduction of the static stability of the aircraft. On the contrast, the static stability of aircraft increases with the increase of aircraft area.

$$K_n = \bar{x}_{c,g} - \bar{x}_{ac} = \frac{\partial C_m}{\partial C_L} \quad (1)$$

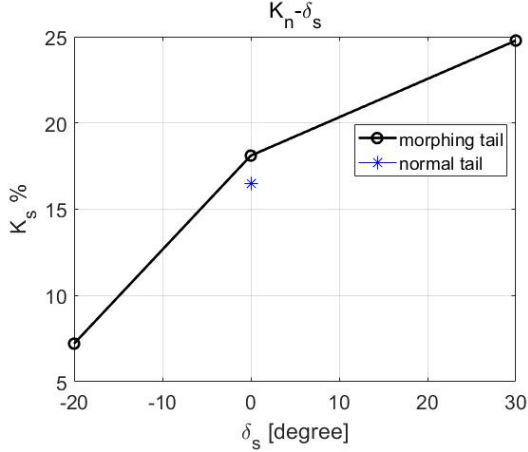

Figure 6. The static stability margin of the aircraft with morphing tail

where $K_n$ is static stability margin of the aircraft, which is used to measure the strength of the longitudinal static stability of an aircraft. $\bar{x}_{ac}$ is the ratio of the aerodynamic center position of the aircraft to the average aerodynamic chord length of the aircraft. $\bar{x}_{c,g}$ is the ratio of the position of the center of gravity of the aircraft to the average aerodynamic chord length. $C_m$ is the longitudinal moment coefficient of the aircraft, $C_L$ is the lift coefficient. Within the actual deformable range from $-20^o$ to $30^o$, the static stability margin increases from 7.21% to 24.75%. Compared with the normal tail (E in Figure. 2), the stability margin of the morphing tail is stronger under the same area of the tail, which is mainly due to the better lift characteristics brought by the airfoil of the morphing tail, which moves the aerodynamic center more backward than the normal tail.

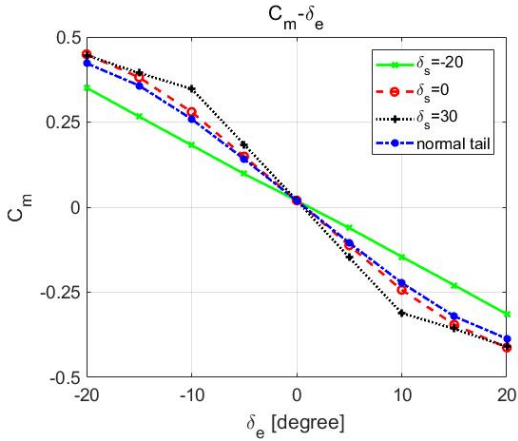

Figure 7. Control efficiency of the morphing tail

With the change of the tail area, the lift coefficient of the morphing tail changes, which leads to the change of the control efficiency of pitch and yaw. For this deformation design, with the increase of $\delta_s$, the sweepback angle of tail gradually decreases. Therefore, the stall characteristics of the morphing tail become worse. For the larger $\delta_s$, the control efficiency is larger in the small $\delta_e$, but it starts to decrease after exceeding a certain $\delta_e$. On the contrary, for the large sweep angle, i.e. the small $\delta_s$, the stall characteristic of the morphing tail is better, so there is a more linear change in the range of control efficiency with the increase of $\delta_e$. Under the same tail projection area, the control efficiency of the deformable tail is similar to that of the conventional tail, but the control efficiency of the morphing tail is slightly better than that of the normal tail.

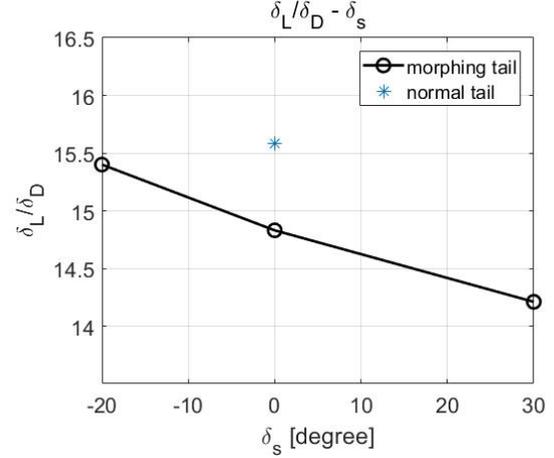

Figure 8. Optimum lif-drag ratio in different $\delta_s$

During the deformation process of the morphing tail, the aerodynamic center of the whole aircraft is changing, the control efficiency of the morphing tail is changing, and the center of gravity of the whole aircraft is also changing. Therefore, the trim control variable of each shape of the morphing tail is different, which will lead to the different lift drag ratio of the whole aircraft. The smaller the tail surface is, the greater the optimum lift-drag ratio will be. $\delta_s$ changing from $-30^o$ to $20^o$, the cruising optimum lift-drag ratio will increase from 14.18 to 15.40, with an increase of 8.37%. Compared with the normal tail, the figure of the morphing tail in same area is 14.83, dropping about 4.81%, but for the best lift-drag ratio 15.40 about the morphing tail, the figure just drops about 1.15%. The morphing tail in this paper is affected by the mechanical structure design, and the minimum $\delta_s$ is only $-20^o$. After the optimization of the mechanism design, it is reasonable to believe that the optimal lift-drag ratio of the morphing tail will be greater than that of the normal tail.

III. MODELING

TABLE III. THE MORPHING AIRCRAFT

| Parameter | Values | Parameter | Values |
|---|---|---|---|
| Mass $m$ | $1.14kg$ | Reference aera $S$ | $0.285m^2$ |
| Wingspan $b$ | $1.40m$ | Reference chord $c$ | $0.211m$ |
| $\delta_a$ | [-25º,25º] | $\delta_e$ | [-20º,20º] |
| $\delta_s$ | [-20º,30º] | $\delta_r$ | [-60º,60º] |

Following [20], the small aircraft can be modeled as a rigid body of fixed mass $m$ in a body frame {b}, moving relative to a NED (north-east-down) frame assumed to be inertial {n}. The aerodynamic data used in the model are calculated by Computational Fluid Dynamics (CFD) and AVL [21], and the aircraft model in this paper is nonlinear. The parameters about the morphing aircraft can be listed as Table Ⅲ. In the modeling, we ignore the change of center of gravity in the process of tail deformation. The aircraft is subject to aerodynamic force, gravity and propeller thrust. The modeling of each part is shown below.

### A. Aerodynamic model

This aircraft is flying in a wind field decomposed into a steady part $v_{w_s}^n$ and a stochastic part $v_{w_g}^n$ representing gusts and turbulence [19]. The steady part is represented in {n}, while the stochastic part is represented in {b}. At the same time, rotational disturbances are modeled through the wind angular velocity $\omega_w$. The attitude of the aircraft can be replaced by unit quaternions $q = [\eta\ \epsilon_1\ \epsilon_2\ \epsilon_3]^T$ where $q^T q = 1$. The relative velocity of the aircraft is the defined as:

$$v_r = v - R_b^n(q)^T v_{\omega_s} - v_{\omega_g} = \begin{bmatrix} u_r \\ v_r \\ w_r \end{bmatrix} \quad (2)$$

$$\omega_r = \omega - \omega_w = \begin{bmatrix} p_r \\ q_r \\ r_r \end{bmatrix} \quad (3)$$

The angle of attack $\alpha$ and sideslip angle $\beta$ can be obtained from the current aircraft speed.

$$V_a = \sqrt{u_r^2 + v_r^2 + w_r^2} \quad (4)$$

$$\alpha = tan^{-1}(\frac{u_r}{w_r}) \quad (5)$$

$$\beta = sin^{-1}(\frac{v_r}{V_a}) \quad (6)$$

The dimension of aerodynamic data is 6 dimensions, such as $C_D(\alpha, \beta, \delta_a, \delta_e, \delta_s, \delta_r)$. According to the current state, the static aerodynamic data and dynamic aerodynamic data of the current state point are obtained by RBF interpolation [22].

$$C_* = C_*(\alpha, \beta, \delta_a, \delta_e, \delta_s, \delta_r) + C_{*p}(\alpha, \beta, \delta_a, \delta_e, \delta_s, \delta_r)\bar{p} + C_{*q}(\alpha, \beta, \delta_a, \delta_e, \delta_s, \delta_r)\bar{q} + C_{*r}(\alpha, \beta, \delta_a, \delta_e, \delta_s, \delta_r)\bar{r} \quad (7)$$

where the $*$ can be D, Y, L, l, m and n. $\bar{p}, \bar{q}, \bar{r}$ are the reference angular velocity.

$$\bar{p} = \frac{pb}{2V_a}, \bar{q} = \frac{qc}{2V_a}, \bar{r} = \frac{rb}{2V_a} \quad (8)$$

where $c$ is the aerodynamic chord, and $b$ is the wingspan of the aircraft. The aerodynamic force and moment of the aircraft are described by

$$F_a = R_w^b \begin{bmatrix} -D \\ Y \\ -L \end{bmatrix} \quad (9)$$

$$\begin{bmatrix} D \\ Y \\ L \end{bmatrix} = \frac{1}{2}\rho V_a^2 S \begin{bmatrix} C_D \\ C_Y \\ C_L \end{bmatrix} \quad (10)$$

$$M_a = \frac{1}{2}\rho V_a^2 S \begin{bmatrix} bC_l \\ cC_m \\ bC_n \end{bmatrix} \quad (11)$$

where $\rho$ is the density of air, $S$ is the reference area of aircraft. The transformation matrix from the wind axis to the body axis, which can be used to transform forces [D Y L].

$$R_w^b = \begin{bmatrix} \cos(\alpha)\cos(\beta) & \cos(\alpha)\sin(\beta) & -\sin(\alpha) \\ -\sin(\beta) & \cos(\beta) & 0 \\ \cos(\beta)\sin(\alpha) & \sin(\alpha)\sin(\beta) & \cos(\alpha) \end{bmatrix} \quad (12)$$

### B. Propulsion Forces and Moments

The motor and propeller are installed on $x_b$ of the body frame, so only thrust and torque are generated in the direction. For the propeller, we carried out the dynamic power experiment, testing the propeller's force, rotating speed and power under different speed $V$ and throttle command $\delta_t$, and fitted the propeller's force and power curve according to the following formula.

$$\left. \begin{aligned} J &= \frac{V}{n_s D_{propeller}} \\ C_T &= \frac{T_t}{\rho n_s^2 D_{propeller}^4} \\ C_p &= \frac{P}{\rho n_s^3 D_{propeller}^5} \end{aligned} \right\} \quad (13)$$

$$M_t = \frac{P}{n_s} \quad (14)$$

where $J$ is the propeller's forward ratio, $C_T$ is the propeller's force coefficient, $C_P$ is the power coefficient. According to the experimental data, we fitted $C_T = f_{C_T}(J)$ and $C_p = f_{C_p}(J)$. $T_t$ and $M_t$ are the force and torque of the propeller, respectively. The speed $V$ refers to the velocity component perpendicular to the propeller disk. $n_s$ is the current propeller speed. The relationship $n_s = f(\delta_t)$ can be fitted through the experimental data. $D_{propeller}$ is the diameter of the propeller disc of the screw propeller, and $D_{propeller} = 0.254m$ in this paper. Therefore, the forces and moments of the power system are:

$$F_{propeller} = \begin{bmatrix} T_t \\ 0 \\ 0 \end{bmatrix}, M_{proeller} = \begin{bmatrix} M_t \\ 0 \\ 0 \end{bmatrix} \quad (15)$$

### C. Actuator Dynamics and Constrains

The model of servo response has an obvious influence on the attitude control effect of the aircraft [20]. Therefore, it is necessary to establish a relatively reasonable model of servo. A second-order integrators is approximately used for the model of servo [23].:

$$\frac{\delta_i}{\delta_i^c} = \frac{\omega_0^2}{s^2 + 2\zeta\omega_0 s + \omega_0^2} \quad (16)$$

for $i = a, e, s, r$, where $c$ is the denoting command, $\omega_0 = 100$ and $\zeta = 0.707$. The throttle dynamic is given as the first order transfer function [24]:

$$\frac{\delta_t}{\delta_t^c} = \frac{1}{Ts+1} \quad (17)$$

where $T = 0.2$.

## IV. CONTROL

There are several reasons for using PPO algorithm in this paper. Firstly, PPO has a best performance on attitude control of quadcopters [17] and fixed-wing aircraft [19]. Secondly, PPO algorithm surpassed the traditional PID algorithm in many indicators. Moreover, the super parameters of PPO algorithm have good robustness for a large variety of tasks.

In this paper, PPO and PID are both used to control the aircraft attitude, including roll, pitch and yaw. There are five control variables ($\delta_t, \delta_a, \delta_e, \delta_s, \delta_r$). $\delta_t$ and $\delta_a$ are controlled by PID, and three control variables ($\delta_e, \delta_s, \delta_r$) about the tail are controlled by PPO. PID is a good model-free control algorithm for single-input single-output system with excellent robustness and accuracy. For roll and speed control of this morphing aircraft, PID is a very suitable control method. For the control of the morphing tail, it is more complex, because the tail has complex control effect and more than one control variables. PPO is suitable for this morphing tail.

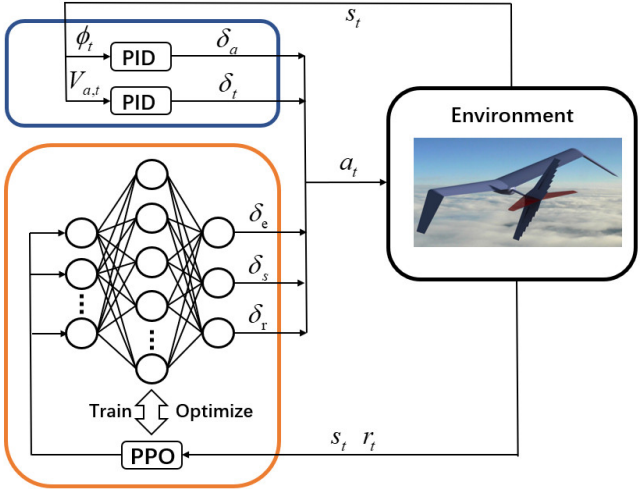

Figure 9. PID&PPO attitude control architecture

### A. PID

PID algorithm in traditional aircraft control has good performance in attitude control. The roll angle and throttle of the aircraft are controlled by this method. Roll angle is controlled by cascade PID:

$$\delta_a = K_{p,P}(p_c - p) + K_{p,I} \int (p_c - p)dt \quad (18)$$

$$p_c = K_\phi(\phi_c - \phi) \quad (19)$$

where $K_{p,P}$ and $K_{p,I}$ are 1.0 and 0 respectively. According to the speed error to control the throttle:

$$\delta_t = K_{t,P}(V_{a,c} - V) + K_{t,I} \int (V_{a,c} - V)dt \quad (20)$$

where $K_{t,P}$ and $K_{t,I}$ are 0.5 and 0.1 respectively, $V$ is the flight speed at the current moment and $V_{a,c}$ is the speed command.

### B. PPO

PPO is a model-free online strategy gradient algorithm. The algorithm combines the advantages of A2C and TRPO. The main feature of the algorithm is that after updating, the new strategy will not change much compared with the old strategy. In order to achieve this goal, PPO algorithm uses the limit to control the number of updates. In this paper, $\pi$ is a strategy function, which is composed of a neural network whose weight parameter is $\theta$ (that is not the pitch angle). Input the current state $s_t$ and output an action $a_t$. For continuous action spaces, the policy network is tasked with outputting the moments of a probability distribution, and the means and variances used in this paper are Gaussian distribution. In the process of training, in order to increase the exploration, the actions are randomly selected from this distribution, while the mean is taken as the action when training is completed. PPO is outlined as following.

---
**Algorithm**: PPO
**for** interation = 1,2, … **do**
    **for** actor = 1,2, … , N **do**
        Run policy $\pi_{old}$ environment for $T$ time steps
        Compute advantage estimates $\hat{A}_1$ , … $\hat{A}_T$
    **end**
    Optimize surrogate $L$ wrt. $\theta$.
    $\theta_{old} \leftarrow \theta$
**end**

---

### C. State space and action space

It is difficult for small UAVs to directly measure the angle of attack and sideslip angle, so these two states are not considered in the observation. The range of observation state is affected by aircraft aerodynamic parameters. For example, too large pitch angle may easily lead to the angle of attack exceeding stall angle of attack.

TABLE IV. STATES RANGE

| Parameter | Initial range | Target range |
|---|---|---|
| $V_a$ | $15\sim 18 m/s$ | $15\sim 18 m/s$ |
| $\phi$ | $-150°\sim 150°$ | $-60°\sim 60°$ |
| $\theta$ | $-60°\sim 60°$ | $-35°\sim 35°$ |
| $\psi$ | $-150°\sim 150°$ | $-60°\sim 60°$ |
| $p$ | $-40\sim 40 deg/s$ | - |
| $q$ | $-40\sim 40 deg/s$ | - |
| $r$ | $-40\sim 40 deg/s$ | - |

The hyperparameters of PPO are tuned wrt. a symmetric action space with a small range from -1 to 1, which is benefit for increasing generality.

### D. Reward Function

The reward function determines the optimization direction of the controller. In the traditional control theory, the general optimization direction is the minimization cost function. In RL algorithm, the optimization direction is the maximum benefit. The reward functions in this paper are all given negative values from -1 to 0. In order to maximize the reward, the agent will converge to the state of less reward loss as soon as possible. The control goal of this paper is to ensure that the attitude quickly reaches the target value, so we need to set the reward function for the attitude. In addition, the state of control variables is also considered in the reward function.

$$\left. \begin{array}{l} R_\phi = clip\left(\frac{|\phi - \phi_{target}|}{\zeta_\phi}, 0, \gamma_\phi\right) \\ R_\theta = clip\left(\frac{|\theta - \theta_{target}|}{\zeta_\theta}, 0, \gamma_\theta\right) \\ R_\psi = clip\left(\frac{|\psi - \psi_{target}|}{\zeta_\psi}, 0, \gamma_\psi\right) \\ R_\delta = clip\left(\frac{\sum_{j\in[a,e,s,r]}\sum_{i=0}^{4}|\delta^c_{j_{t-1}} - \delta^c_{j_{t-1-i}}|}{\zeta_\delta}, 0, \gamma_\delta\right) \end{array} \right\} \quad (21)$$

$$R_t = -(R_\phi + R_\theta + R_\psi + R_\delta) \quad (22)$$

$$\zeta_\phi = 2.62, \zeta_\theta = 2.04, \zeta_\psi = 2.62, \zeta_\delta = 60 \quad (23)$$

$$\gamma_\phi = 0.30, \gamma_\theta = 0.30, \gamma_\psi = 0.30, \gamma_\delta = 0.1 \quad (24)$$

In this reward function, $R_\phi$, $R_\theta$ and $R_\psi$ are the reward parts for attitude, which compress the reward value between 0 and 0.3. The function of $\zeta$ is to set the reward threshold of attitude. The threshold values of roll angle, pitch angle and yaw angle are 45, 35 and 45 degrees respectively. The reward is 0.3 when the absolute value of the difference between the current attitude and the target attitude is outside the thresholds. The smaller the difference between the absolute value, that is, the closer to the target value, the smaller the reward we get. Because the final reward is negative, it means that the closer to the absolute value, the smaller the punishment. In the optimal control, it is not uncommon that the control command switching between the maximum and the minimum. Such extreme changes have a great impact on the aerodynamic efficiency of the aircraft and the durability of the control devices. Therefore, it is necessary to add a reward function for the control variables. If the control variables continuously change states very big in a short time, it will give the penalty, which can promote the controller in the training process does not change frequently.

## V. TRAINING AND RESULT

In general, the sampling frequency of flight control system of small UAV is 100Hz, so the simulation time step in this paper is 0.01s. The training environment inherits GYM [25]. The controller was trained on a desktop computer with an Inter(R) Xeon(R) Gold 5118 CPU, without a GPU. After 30 parallel environments at the same time, 5 million steps of training were carried out, which lasted for about 5 days.

### A. Key Factors Impacting Training

The selection of state variables has a great impact on training. This paper finds that the current attitude angle and attitude angular acceleration are the most important factors for attitude angle control. When these state variables are not observed, the result is relatively bad. For example, when the roll angle acceleration is reduced, the aircraft will have attitude acceleration oscillation after stabilization.

At the same time, the setting of reward function has a great impact on the control effect of the aircraft. In RL, reward engineering is the process of designing a reward system to provide the agent a signal showing that they are doing the right thing [26]. In this paper, the roll angle is controlled by PID, but the reward function of roll angle should also be considered in PPO. If not, the attitude of the aircraft cannot be controlled very well, and the roll angle is in a state of constant oscillation. Because the RL control model assumes that roll angles do not have a significant effect on getting the best reward.

### B. Evaluation and Result

In the training, each episode is set a random setpoint of attitude and speed. After training, we set a new setpoint, and the controller is perfectly capable of adapting to this new setpoint. The controller is trained in the simulator without wind estimates, but we test the trained controller with different turbulence in the test environment. This test was also done by Koch et al. [17] and Eivind et al. [19].

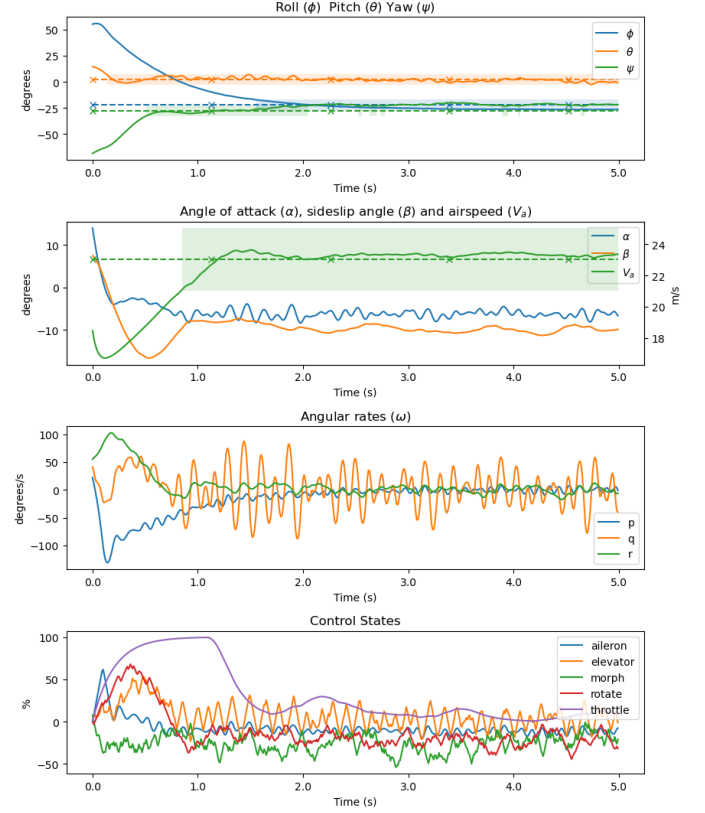

Figure 10. Controller was tested with a new random setpoint

Table V shows the performance of trained controller in some criterions, including success rate, rise time, overshoot and stability time. The success rate represents the success probability of the aircraft to reach the specified attitude target in different initial states. The success rate of the controller is evaluated by 200 complete flight processes. The rise time is defined in this paper as the time interval from 10% to 90% of the initial error. In this paper, the ratio of the error corresponding to the maximum overshoot point to the initial error is used to represent the overshoot of the controller. The stability time represents the time when each attitude is within 5% of the target value.

TABLE V. PERFORMANCE METRICS FOR THE CONTROLLER ON THE EVALUATION SCENARIOS

| Turbulence Setting | Success (%) | | | | Rise time (s) | | | | Settling time (s) | | | | Overshoot (%) | | | |
|---|---|---|---|---|---|---|---|---|---|---|---|---|---|---|---|---|
| | $\phi$ | $\theta$ | $\psi$ | $V_a$ | $\phi$ | $\theta$ | $\psi$ | $V_a$ | $\phi$ | $\theta$ | $\psi$ | $V_a$ | $\phi$ | $\theta$ | $\psi$ | $V_a$ |
| No | 100 | 100 | `100 | 100 | 1.49 | 0.72 | 2.62 | 1.09 | 2.87 | 1.76 | 5.34 | 1.82 | 5.42 | 25.6 | 31.2 | 21.1 |
| Light | 100 | 99 | 98 | 96 | 0.94 | 1.01 | 3.42 | 0.94 | 3.01 | 1.82 | 6.78 | 2.03 | 20.6 | 19.3 | 45.2 | 39.6 |
| Moderate | 90 | 95 | 73 | 87 | 1.70 | 1.11 | 3.06 | 0.41 | 3.58 | 2.03 | 5.62 | 1.69 | 10.7 | 27.9 | 153 | 91 |
| Severe | 81 | 82 | 65 | 79 | 0.81 | 0.78 | 1.95 | 0.43 | 3.67 | 2.16 | 5.90 | 2.05 | 40.2 | 30.7 | 83.9 | 106 |

## VI. Conclusion

The bionic tail with the multi-dimensional control has excellent performance in attitude control. Birds use multi-dimensional tail to ensure the stability in a variety of complex states. In this paper, the tail is used to control the attitude, including yaw angle. In the fixed wing aircraft without vertical tail, the aircraft adopts the joint control of aileron and elevator to control the yaw angle. After using the bionic tail, the aircraft can control the yaw angle by using the tail with only one control surface. The yaw angle is controlled by the rotation of the body shafting. I think the morphing tail can be used for other purposes, such as balancing the yaw angle changes caused by aircraft roll control, or used to eliminate the sideslip angle to improve the flight performance of the aircraft, and to improve the stability of the aircraft. This is closely related to the control objectives of the tail. In this paper, we only use the tail to control the attitude of the aircraft. The bionic tail has some obvious advantages, including adjusting the stability and the maneuverability of the aircraft. Changing the lift-drag characteristics of the aircraft is another advantage. When the tail area becomes smaller, the lift-drag ratio of the aircraft is improved to a certain extent.

In this paper, the model free DRL control algorithm is used to avoid the problems caused by the highly nonlinear aircraft model. The controller is optimized by continuous interaction with the environment. Finally, it has good performance in attitude control. Reinforcement learning has a good performance in multi input and multi output control objects, which sometimes are difficult to establish accurate model. At the same time, there are some problems to be solved, such as the difference between the simulation environment and the real environment, whether the controller trained in the simulation environment is safe enough to use in the real environment, especially in the field of aircraft control, where a simple failure may cause damage to the aircraft and cost a lot of money and time. DRL has the problem of over fitting. How to improve the robustness of the controller in flight control is also a problem to be solved. In addition, reward engineering is also a challenge to reach the desired result.